\documentclass
[superscriptaddress,secnumarabic,amssymb,amsmath,nobibnotes,aps,prd,showkeys,showpacs,nofootinbib,onecolumn]{revtex4}%
\usepackage{graphicx}
\usepackage{epstopdf}
\usepackage{epsf}
\usepackage{bm}
\usepackage{amsmath}
\usepackage{amsfonts}
\usepackage{amssymb}%
\providecommand{\U}[1]{\protect\rule{.1in}{.1in}}

\newcommand{\be}{\begin{equation}}
\newcommand{\ee}{\end{equation}}

\newcommand{\mincir}{\raise
-3.truept\hbox{\rlap{\hbox{$\sim$}}\raise4.truept\hbox{$<$}\ }}
\newcommand{\magcir}{\raise
-3.truept\hbox{\rlap{\hbox{$\sim$}}\raise4.truept\hbox{$>$}\ }}

\begin{document}
\title{Cosmological Evolution and Exact Solutions in a Fourth-order Theory of Gravity}
\author{Andronikos Paliathanasis}
\email{anpaliat@phys.uoa.gr}
\affiliation{Instituto de Ciencias F\'{\i}sicas y Matem\'{a}ticas, Universidad Austral de
Chile, Valdivia, Chile}
\affiliation{Institute of Systems Science, Durban University of Technology, Durban 4000,
South Africa}

\begin{abstract}
A fourth-order theory of gravity is considered which in terms of dynamics has
the same degrees of freedom and number of constraints as those of
scalar-tensor theories. In addition it admits a canonical point-like
Lagrangian description. We study the critical points of the theory and we show
that it can describe the matter epoch of the universe and that two accelerated
phases can be recovered one of which describes a de Sitter universe. Finally
for some models exact solutions are presented.

\end{abstract}
\date{\today}
\maketitle

\section{Introduction}

The origin of the late-time acceleration phase of the universe is an unsolved
puzzle in modern cosmology \cite{Teg,Kowal,Komatsu,Ade15}. The effects of the
late-time acceleration have been attributed to the so-called dark energy. As
dark energy is characterized as the matter source which provides the missing
terms in the field equations of Einstein's General Relativity and which leads
to solutions that describe the accelerating expansion of the universe. The
proposed solutions for the nature of dark energy can be categorized into two
big classes: (i) the dark energy models where an energy momentum tensor, which
describes an exotic matter source
\cite{Ratra,Barrow,Linder,Overduin,Bento,Kame,Gal1,Gal2,Supri}, is introduced
into Einstein's General Relativity and/or (ii) the Einstein-Hilbert action is
modified such that the new field equations provide additional terms which are
assumed to contribute to the acceleration of the universe; for instance see
\cite{Brans,Buda,Sotiriou,odin1,cap01,Ferraro,Ferraro2,basStav,kof1,myr,od4,od5,ma1,ma2}%
. In the second approach, the dark energy has a geometrical origin and
description \cite{capQ}.\ Some recent cosmological constraints of modified
theories of gravity are given in \cite{Ade15b,data1,data2,nunes1,nunes2} while
some solar system tests can be found in \cite{sol1,sol2,sol3,sol4}.

A special class of modified theories of gravity which has drawn attention over
recent years comprises the $f\left(  X\right)  $ theories of gravity where $X$
is an invariant, for instance the Ricci Scalar, $R$, of the underlying space.
In the latter case the theory is the well-studied $f\left(  R\right)
$-gravity \cite{Buda} which in general is a fourth-order gravitational theory.
Furthermore, when it is a second-order theory General Relativity is recovered
because function $f$ ~has to be a linear function. Though the functional form
of the theory which describes the universe it is unknown, however, various
(toy) models have been proposed in the literature in order to describe
different phenomena, for a review see \cite{Sotiriou}. As we mentioned above,
$f\left(  R\right)  $-gravity is a fourth-theory and the dependent variables
of the gravitational field equations are the one which follow from the line
element which defines the spacetime.

Like every fourth-order differential equation which can be written as a system
of a two second-order differential equation, $f\left(  R\right)  $-gravity can
be written as a second-order theory by introducing a new degree of freedom.
That new degree of freedom it is equivalent with that of Brans-Dicke scalar
field, with zero Brans-Dicke parameter. The equivalence of a modified theory
with a scalar field it is not true for all the $f\left(  X\right)  $-theories,
but always it depends upon the nature of the invariant(s) $X$.

Another $f\ $theory of special interest is the $f\left(  T\right)  $
teleparallel gravity \cite{Ferraro} in which $T$ is related to the
antisymmetric connection used in the theory \cite{Hayashi79}. In the $f\left(
T\right)  $-gravity General Relativity, and specifically the teleparallel
equivalence of General Relativity, is recovered when the function $f$ is
linear \cite{ein28}. Because $T$ admits terms with first derivatives,
$f\left(  T\right)  $-gravity provides a theory in which the gravitational
field equations are of second-order. However, $f\left(  T\right)  $-gravity in
general provides different properties from General Relativity
\cite{ftSot,ftTam}. Moreover there is not any scalar-field/scalar tensor
description like in $f\left(  R\right)  $-gravity and in terms of dynamics the
terms which gives the dark energy description can be seen as extra constraints
on the dynamical system.

The existence of other invariants in the Action Integral provides different
components in the modified gravitational field equations. Some other theories
which have been proposed are the modified $R+f\left(  G\right)  $ Gauss-Bonnet
gravity \cite{mod1,on2}, the more general $f\left(  R,G\right)  $ Gauss Bonnet
gravity \cite{on3,mod2}, the $f\left(  R,T^{\left(  m\right)  }\right)
$-gravity, where $T^{\left(  m\right)  }$ is the trace of the energy momentum
tensor \cite{mod3}, and many others. In this work we are interested in the
so-called $f\left(  R,T\right)  $ gravity \cite{myr11}, where $R$ is the Ricci
Scalar of the underlying space and $T$ the invariant of teleparallel gravity.
That theory is equivalent with the proposed $f\left(  T,B\right)  $ theory
\cite{bahamonde}, where $B$ is the boundary term which relates $T$ and $R$,
and specifically~$B=2e_{\nu}^{-1}\partial_{\nu}\left(  eT_{\rho}^{~\rho\nu
}\right)  $ so that $R=-T+B,~$where $T_{\mu\nu}^{\beta}$ is the curvatureless
Weitzenb\"{o}ck connection.

For the dynamics of the field equations it is easy to see that $f\left(
R,T\right)  $ is a fourth-order theory of gravity, but new constraints follow
from the $T$ terms. However, in the limit in which $f_{,TT}=0$ the theory is
reduced to that of $f\left(  R\right)  $-gravity and has the same number of
constraints as the Brans-Dicke theory. However, that is not the unique case in
which the theory has the same number of constraints, in the dynamics, as that
of scalar-tensor theories. As we see below that property exists and for\ the
$f\left(  R,T\right)  \equiv T+F\left(  R+T\right)  ~$theory, or in the
equivalent description, for the $f\left(  T,B\right)  \equiv T+F\left(
B\right)  $ theory of gravity ~or $f\left(  R,B\right)  =R+F\left(  B\right)
$.~This is the toy-model that we study in this work. The plan of the paper is
as follows.

In Section \ref{field} we define our cosmological model and with the use of a
Lagrange multiplier we derive the gravitational field equations. In order to
study the general evolution of the field equations in Section \ref{cosmevol}
we study the critical points for an arbitrary function $F\left(  R+T\right)
$. We find it is possible for the theory to provide two accelerated eras, one
stable and one unstable, which can be related with the early acceleration
phase (inflation) and the late acceleration phase. Moreover we see that there
exists a critical point in which the scalar field mimics the additional
perfect fluid that we consider exists and that point can describe the
matter-dominated epoch of the universe. Furthermore, some closed-form
analytical solutions are presented in Section \ref{solutions} while in Section
\ref{conclusions} we draw our conclusions and discuss our results.

\section{The field equations}

\label{field}

The theory that we are interesting is a special form of $f\left(  R,T\right)
$-gravity in which $R$ is the Ricci Scalar of the underlying space and $T$ the
the invariant of Weitzenb\"{o}ck connection. The two quantities are related by
the expression $R=-T+B$,~where
\begin{equation}
B=2e_{\nu}^{-1}\partial_{\nu}\left(  eT_{\rho}^{~\rho\nu}\right)
\end{equation}
is the boundary term. We follow the notation of \cite{bahamonde} and we find
that for $f\left(  T,R+T\right)  =f\left(  T,B\right)  $ gravity the
gravitational field equations are%
\begin{align}
16\pi Ge\mathcal{T}_{a}{}^{\lambda}  &  =2eh_{a}^{\lambda}\left(
f_{,B}\right)  ^{;\mu\nu}g_{\mu\nu}-2eh_{a}^{\sigma}\left(  f_{,B}\right)
_{;\sigma}^{~~~;\lambda}+eBh_{a}^{\lambda}f_{,B}\,+4(eS_{a}{}^{\mu\lambda
})_{,\mu}f_{T}\nonumber\\
&  ~\ ~+4e\Big[(f_{,B})_{,\mu}+(f_{,T})_{,\mu}\Big]S_{a}{}^{\mu\lambda
}~-4ef_{,T}T^{\sigma}{}_{\mu a}S_{\sigma}{}^{\lambda\mu}-efh_{a}^{\lambda},
\label{fe.02}%
\end{align}
where {$\mathcal{T}_{\rho}{}^{\nu}$} is the energy-momentum tensor of the
matter source, a comma denotes partial derivative, \textquotedblleft%
$;$\textquotedblright\ $\ $ denotes covariant derivative and $e_{i}=h_{i}%
^{\mu}\left(  x\right)  \partial_{i}$ is the vierbein field which defines
the~Weitzenb\"{o}ck connection,~$\hat{\Gamma}^{\lambda}{}_{\mu\nu}%
=h_{a}^{\lambda}\partial_{\mu}h_{\nu}^{a}$, where~\ $T_{\mu\nu}^{\beta}%
=\hat{\Gamma}_{\nu\mu}^{\beta}-\hat{\Gamma}_{\mu\nu}^{\beta}=h_{i}^{\beta
}(\partial_{\mu}h_{\nu}^{a}-\partial_{\nu}h_{\mu}^{a}).~$Moreover~${S_{\beta}%
}^{\mu\nu}=\frac{1}{2}({K^{\mu\nu}}_{\beta}+\delta_{\beta}^{\mu}{T^{\theta\nu
}}_{\theta}-\delta_{\beta}^{\nu}{T^{\theta\mu}}_{\theta})\,~$and ${K^{\mu\nu}%
}_{\beta}$ is the cotorsion tensor given by the expression
\begin{equation}
{K^{\mu\nu}}_{\beta}=-\frac{1}{2}({T^{\mu\nu}}_{\beta}-{T^{\nu\mu}}_{\beta
}-{T_{\beta}}^{\mu\nu}) \label{ft.04}%
\end{equation}
and equals the difference between the Levi-Civita connections in the holonomic
and the nonholonomic frame\footnote{For more details on the covariant
formulation of teleparallel gravity we refer the reader to \cite{ss12}}.
Finally $e=\det(e_{\mu}^{i})=\sqrt{-g}$.

For the gravitational field equations (\ref{fe.02}) it is easy to see that,
when $f_{,BB}=0$, the field equations reduce to those of $f\left(  T\right)  $
teleparallel gravity while, as it has been mentioned in \cite{bahamonde}%
,$~$for $f\left(  T,B\right)  =f\left(  -T+B\right)  $, $f\left(  R\right)
$-gravity is recovered. Last but not least in general the field equations
(\ref{fe.02}) are of fourth-order.

We assume that the geometry which describe the universe is that of a spatially
flat Friedmann-Lema\^{\i}tre-Robertson-Walker (FLRW) spacetime with line
element%
\begin{equation}
ds^{2}=-N(t)^{2}dt^{2}+a(t)^{2}\left(  dx^{2}+dy^{2}+dz^{2}\right)  ,
\label{s0001}%
\end{equation}
where $a\left(  t\right)  $ is the scale factor and $N\left(  t\right)  $ is
the lapse function. Moreover we consider the diagonal frame for the vierbein
to be,
\begin{equation}
h_{\mu}^{i}(t)=diag(N\left(  t\right)  ,a(t),a(t),a(t)) \label{ft.07}%
\end{equation}
from which we calculate that
\begin{equation}
T=-\frac{6}{N^{2}}\left(  \frac{\dot{a}}{a}\right)  ^{2}~,~B=-\frac{6}{N^{2}%
}\left(  \frac{\ddot{a}}{a}+\frac{2\dot{a}^{2}}{a^{2}}-\frac{\dot{a}\dot{N}%
}{aN}\right)  . \, \label{ft.002}%
\end{equation}

For that frame and for the comoving observer, $u^{\lambda}=N^{-1}\delta
_{0}^{t}$, $\left(  u^{\lambda}u_{\lambda}=-1\right)  $, the gravitational
field equations are
\begin{equation}
\frac{f}{2}-\frac{3\dot{a}\dot{f}_{,B}}{aN^{2}}+\frac{6f_{T}\dot{a}^{2}}%
{a^{2}N^{2}}+\frac{3f_{,B}}{N^{2}}\left(  \frac{\ddot{a}}{a}-\frac{\dot{a}%
\dot{N}}{aN}+2\left(  \frac{\dot{a}}{a}\right)  ^{2}\right)  =\rho\label{s01}%
\end{equation}
and
\begin{equation}
\frac{f}{2}+\frac{2\dot{a}\dot{f}_{,T}}{aN^{2}}+\frac{\left(  3f_{,B}%
+2f_{,T}\right)  }{N^{2}}\left(  \frac{\ddot{a}}{a}-\frac{\dot{a}\dot{N}}%
{aN}+2\left(  \frac{\dot{a}}{a}\right)  ^{2}\right)  -\frac{\ddot{f}_{,B}%
}{N^{2}}+\frac{\dot{f}_{,B}\dot{N}}{N^{3}}=-p\, \label{s02}%
\end{equation}
where overdot denotes total derivative with respect to $t$~and $\left\{
\rho,~p\right\}  $ are the energy density, $\rho=\mathcal{T}_{\mu\nu}u^{\mu
}u^{\nu}$ and the pressure~$p=\mathcal{T}_{\mu\nu}\left(  g^{\mu\nu}+u^{\mu
}u^{\nu}\right)  $ of the matter source.

\subsection{Lagrange multiplier and minisuperspace}

As we have already mentioned, $f\left(  T,B\right)  $-gravity is a
fourth-order theory. However, \ as in the case of $f\left(  R\right)
$-gravity Lagrange multipliers can be introduced in order to reduce the order
of the differential equations. However, the latter means that the degrees of
freedom are increased. \ Therefore from the definition of $T$ and $B$, that is
expression (\ref{ft.002}), and with the introduction of the Lagrange
multipliers, $\lambda_{1}$ and $\lambda_{2}$, the gravitational Action
Integral becomes
\begin{equation}
A=\int dt\left[  fNa^{3}-\lambda_{1}\left(  T+6\left(  \frac{\dot{a}}%
{Na}\right)  ^{2}\right)  -\lambda_{2}\left(  B+\frac{6}{N^{2}}\left(
\frac{\ddot{a}}{a}+\frac{2\dot{a}^{2}}{a^{2}}-\frac{\dot{a}\dot{N}}%
{aN}\right)  \right)  \right]  , \label{actioncan}%
\end{equation}
where for simplicity we have assumed the vacuum case.

Variation of the Action Integral above with respect to the variables, $T$ and
$B$, provides the definition of $\lambda_{1}$ and $\lambda_{2}~$from the
expression~$\frac{\delta A}{\delta T}=0~,~\frac{\delta A}{\delta B}%
=0.~$Therefore we find that%
\[
\lambda_{1}=Na^{3}f_{,T}~\text{\ and~~}\lambda_{2}=Na^{3}f_{,B}.
\]

Hence the gravitational action becomes%
\begin{equation}
A=\int dt\left[  (fNa^{3}-Na^{3}f_{,T}\left(  T+6\left(  \frac{\dot{a}}%
{Na}\right)  ^{2}\right)  -Na^{3}f_{,B}\left(  B+\frac{6}{N^{2}}\left(
\frac{\ddot{a}}{a}+\frac{2\dot{a}^{2}}{a^{2}}-\frac{\dot{a}\dot{N}}%
{aN}\right)  \right)  \right]  , \label{actioncan2}%
\end{equation}
from which by integration by parts we find the Lagrangian of the field
equations to be%
\begin{equation}
\mathcal{L}_{f(T,B)}=-\frac{6}{N}a\dot{a}^{2}f_{,T}+\frac{6}{N}a^{2}\dot
{a}\dot{f}_{,B}+Na^{3}\left(  f-Tf_{,T}-Bf_{,B}\right)  . \label{L}%
\end{equation}
Finally the field equations are given from the Euler-Lagrange equations of
(\ref{L}) with respect to the variables $\left\{  N,a,T,B\right\}  $, where
$\frac{\partial L}{\partial N}=0$, is the constraint equation.

Without loss of generality we can assume the lapse function to be $N\left(
t\right)  =1$. We define the new variable $\phi=f_{,B}$. Thus the Lagrangian
(\ref{L}) takes the simpler form%
\begin{equation}
\mathcal{L}_{f(T,B)}=-6a\dot{a}^{2}f_{T}+6a^{2}\dot{a}\dot{\phi}%
\,-a^{3}V\left(  \phi,T\right)  , \label{L.01}%
\end{equation}
where now%
\begin{equation}
V\left(  \phi,T\right)  =Tf_{,T}+Bf_{,B}-f\left(  T,B\right)  \text{.}
\label{L.02}%
\end{equation}

The field equations in $f\left(  T,B\right)  $-gravity are in general of
fourth-order, except when $f_{B}$ is constant. By introducing the field $\phi
$, the Lagrangian (\ref{L.01}) describes the evolution of a dynamical system
in the space of variables $\left\{  a,\phi,T\right\}  $ while, when
$f=f\left(  T-B\right)  $, we see that $f_{,T}=-f_{,B}=-\phi$, which means
that the Lagrangian of O'Hanlon gravity\footnote{The Action Integral of the
O'Hanlon theory it coincides with that of Brans-Dicke theory for zero
Brans-Dicke parameter. However, the theory has been introduced in order to
produce a Yukawa type interaction in the gravitational potential \cite{Wand}.}
\cite{Hanlon} is recovered.

Furthermore it is easy to see that (\ref{L.01}) is a singular Lagrangian when
$f_{TT}\neq0$, as the Lagrangian of the field equations is in $f\left(
T\right)  $-gravity.

\subsubsection{Field equations in$~f\left(  T,B\right)  =T+F\left(  B\right)
$}

Inspired from the other modified theories of gravity, specifically from
$f\left(  R\right)  $, for which models of the form $f\left(  R\right)
=R+F\left(  R\right)  $, or in $f\left(  T\right)  $ with $f\left(  T\right)
=T+F\left(  T\right)  $ have been proposed \cite{Ferraro}, here we select to
work with the theory $f\left(  T,B\right)  =T+F\left(  B\right)  $, which is
exactly equivalent to the theories $f\left(  R,B\right)  =R+F\left(  B\right)
$~or $f\left(  R,T\right)  =T+F\left(  T+R\right)  $.

The main characteristic of that selection is that the Lagrangian of the field
equations\ (\ref{L.01}) is a regular Lagrangian in the space of variables
$\left\{  a,\phi\right\}  $, as also is independent of $T.$ Moreover for small
values of the function, $F\left(  B\right)  $, we are in a small deviation
from General Relativity while, for $F_{,B}\left(  B\right)  =0$, General
Relativity is recovered.

In our cosmological scenario we consider a perfect fluid with constant
equation of state parameter $p_{m}=w\rho_{m}$. For $f=T+F(B)$ the
gravitational field equations are%
\begin{equation}
3H^{2}-3H\dot{\phi}-\frac{1}{2}V\left(  \phi\right)  =\rho_{m}, \label{L.03}%
\end{equation}

\begin{equation}
\dot{H}+3H^{2}+\frac{1}{6}V_{,\phi}=0, \label{L.04}%
\end{equation}
and
\begin{equation}
\ddot{\phi}+3H^{2}+\frac{1}{2}V+\frac{1}{3}V_{,\phi}-p_{m}=0. \label{L.05}%
\end{equation}
Moreover, we assume that the there is not any interaction in the Action
integral of the matter source with the gravitation terms the Bianchi identity
provides {the conservation equation\footnote{The lhs of equations
(\ref{L.03})-(\ref{L.05}) can be calculated easily from (\ref{s01}%
)-(\ref{s02}) by assuming $\phi=f_{,B}$, or from the action of the
Euler-Lagrange operator on the Lagrangian (\ref{L.02}).}}%
\begin{equation}
\dot{\rho}_{m}+3\left(  \rho_{m}+p_{m}\right)  H=0\,. \label{L.06}%
\end{equation}

We observe that (\ref{L.03}) is the constrain equation (\ref{s01}), while
equation (\ref{L.04}) describes the evolution of the Hubble function. The
third equation (\ref{L.05}) is the \textquotedblleft
Klein-Gordon\textquotedblright\ (-like) equation for the field $\phi$~which
with the use of (\ref{L.04}) gives the fourth-order equation. However while
the fourth-order $f\left(  R\right)  $-gravity is equivalent with a
Brans-Dicke scalar field, and specifically with the O'Hanlon theory
\cite{Hanlon}, that is not true for our model where indeed the higher-order
derivatives are describing by the field $\phi$, but it is not a canonical
field. On the other hand the field equations are more close to that of a
particle in the Generalized Uncertainty principle~\cite{Supri}, where as the
position of the particle now we consider that of the scale factor $a(t)$.

In the following sections we study the general evolution of the field
equations (\ref{L.03})-(\ref{L.06}) and we search for analytical solutions of
the field equations for specific forms of $V\left(  \phi\right)  $. Recall
that now the partial differential equation (\ref{L.02}) has been reduced to
the Clairaut first-order differential equation
\begin{equation}
V\left(  F_{B}\right)  =BF_{,B}-F\left(  B\right)  \,. \label{L.06a}%
\end{equation}

The latter has always two solutions, the linear solution $F_{S}\left(
B\right)  =F_{1}B+F\left(  F_{1}\right)  $, for arbitrary potential $V\left(
\phi\right)  $, as also a singular solution which is given from the solution
of the second-order differential equation $\frac{dV\left(  F_{B}\right)
}{d\left(  F_{B}\right)  }-B=0.~$The latter solution is the one in which we
are interested, because for the linear solution $F_{S}$ we are in General
Relativity in which $F\left(  F_{1}\right)  $ plays the role of the
cosmological constant.

We continue with the study of the dynamics of the field equations
(\ref{L.03})-(\ref{L.06}). \qquad

\section{Cosmological evolution}

\label{cosmevol}

In order to perform our analysis we assume that the matter source, $\rho
_{m},~p_{m}$, is a perfect fluid with constant equation of state parameter
$w_{m}$, i.e., $p_{m}=w_{m}\rho_{m}$. We define the new dimensionless
variables%
\begin{equation}
x=\frac{\dot{\phi}}{H}~\ ,~y=\frac{1}{6}\frac{V\left(  \phi\right)  }{H^{2}%
}~,~\Omega_{m}=\frac{\rho_{m}}{3H^{2}} \label{L.07}%
\end{equation}
in analogue to the scalar tensor theories \cite{copeland,dn2,dn3,dn4}.
Equation (\ref{L.03}) provides us with the constraint equation%
\begin{equation}
\Omega_{m}=1-x-y \label{L.08}%
\end{equation}
which holds for value of $\left(  x,y\right)  $ where $0\leq\Omega_{m}\leq1$.

We define the new lapse, $N=\ln a$, and now in the new variables the field
equations form the following system of first-order differential equations
\begin{equation}
\frac{dx}{dN}=-3\left(  1+y-x\right)  +\lambda y\left(  2-x\right)
+3w_{m}\Omega_{m}, \label{L.09}%
\end{equation}%
\begin{equation}
\frac{dy}{dN}=\left(  6-\lambda\left(  2y+x\right)  \right)  y \label{L.10}%
\end{equation}
and
\begin{equation}
\frac{d\lambda}{dN}=-x\lambda^{2}\bar{\Gamma}\left(  \lambda\right)  ,
\label{L.11}%
\end{equation}
where
\begin{equation}
\lambda=-\frac{V_{,\phi}}{V}~,~\bar{\Gamma}\left(  \lambda\right)
=\frac{V_{,\phi\phi}}{\left(  V_{,\phi}\right)  ^{2}}-1. \label{L.12}%
\end{equation}

Finally the equation of state parameter for the total fluid, $w_{tot}%
=-1-\frac{2}{3}\frac{\dot{H}}{H^{2}}$, is expressed as a function of
$x,y,\lambda$ as follows
\begin{equation}
w_{tot}=1-\frac{2}{3}\lambda y. \label{L.13}%
\end{equation}

We study the general evolution of the system (\ref{L.09})-(\ref{L.11}) for the
two cases: (a) $\lambda=const$, which means $\bar{\Gamma}\left(
\lambda\right)  =0$, i.e.~$V\left(  \phi\right)  =V_{0}e^{-\lambda\phi}$ and
(b) $\lambda\neq0$. In case (a) the system (\ref{L.09})-(\ref{L.11}) reduces
to a two-dimensional system.

\subsection{Exponential potential}

Consider now that $V\left(  \phi\right)  =V_{0}e^{-\lambda\phi}$, which
corresponds to the function
\begin{equation}
F\left(  B\right)  =-\frac{B}{\lambda}\left(  \ln\left(  -\frac{B}{\lambda
}\right)  -1\right)  .
\end{equation}

For that potential, the fixed points of the dynamical system, (\ref{L.09}%
)-(\ref{L.10}), are the points: $P_{A}~$where $\left(  x_{A},y_{A}\right)
=\left(  1,0\right)  $, $P_{B}$ with~$\left(  x_{B},y_{B}\right)  =\left(
\frac{3}{\lambda}\left(  1+w\right)  ,\frac{3}{2\lambda}\left(  1-w\right)
\right)  $ and $P_{C}$ with coordinates $\left(  x_{C},y_{C}\right)  =\left(
-\frac{6}{\lambda}+2,\frac{6}{\lambda}-1\right)  $. \ Specifically for each
point we have:

\begin{itemize}
\item Point $P_{A}$ corresponds to a universe without matter source;
$\Omega_{m}=0$ and the total equation of state parameter is $w_{tot}=1$, that
is, the scalar field behaves like a stiff matter. The eigenvalues of the
linearized system are calculated to be
\begin{equation}
e_{1}^{A}=3\left(  1-w_{m}\right)  ,~e_{2}^{A}=6-\lambda.
\end{equation}
For the range $w_{m}\in\lbrack-1,1)$, the eigenvalue $e_{1}^{A}$ is always
positive. Hence the point is unstable.

\item At point $P_{B}$ the physical quantities are calculated to be
$\Omega_{m}=1-\frac{3\left(  3+w_{m}\right)  }{2\lambda}$ and $w_{tot}=w_{m}$.
The point is well defined for $\lambda\geq\frac{\left(  9+3w\right)  }{2}$. In
that point the field $\phi$ mimics the matter source as the analogy of the
exponential model in the minimally coupled cosmological scenario, which
describes an accelerated universe for $w_{m}<-\frac{1}{3}$. \newline The two
eigenvalues of the linearized system are%
\begin{align}
e_{1}^{B}  &  =\frac{1}{4}\left(  -3\left(  1-w_{m}\right)  -\sqrt{3}%
\sqrt{\left(  1-w_{m}\right)  \left(  75+21w_{m}-16\lambda\right)  }\right)
,\\
e_{2}^{B}  &  =\frac{1}{4}\left(  -3\left(  1-w_{m}\right)  +\sqrt{3}%
\sqrt{\left(  1-w_{m}\right)  \left(  75+21w_{m}-16\lambda\right)  }\right)  .
\end{align}
For values of $w_{m}$ in the range $w_{m}\in\lbrack-1,1)$, $\mathit{Re}\left(
e_{1}^{B}\right)  <0$ always holds. Furthermore for $\lambda\geq\frac{3}%
{16}\left(  25+7w_{m}\right)  ,~\lambda_{2}^{B}=\frac{3}{16}\left(
25+7w_{m}\right)  $, $\mathit{Re}\left(  e_{2}^{B}\right)  =\mathit{Re}\left(
e_{1}^{B}\right)  <0$ holds, which means that the point is stable. However, if
$\lambda<\lambda_{2}^{B}$, the point is stable when $\lambda_{1}^{B}%
<\lambda<\lambda_{2}^{B}$~in which $\lambda_{1}^{B}=\frac{3}{2}\left(
3+w_{m}\right)  $. Hence we conclude for $\lambda>\lambda_{1}^{B}$ the point
$P_{B}$ is always stable.

\item Point $P_{C}$ describes a universe dominated by the field $\phi$, where
$\Omega_{m}=0$ and equation of state for the total fluid is $w_{tot}=\frac
{1}{3}\left(  2\lambda-9\right)  $, which describes an accelerated universe
for $\lambda<\frac{9}{2}$ and describes a de Sitter universe for $\lambda=3~$.
The eigenvalues of the linearized system are%
\begin{equation}
e_{1}^{C}=-6+\lambda~,~e_{2}^{C}=-9-3w_{m}+2\lambda
\end{equation}
which gives that point is stable when $\lambda<6$ and $\lambda<\frac{3}%
{2}\left(  3+w_{m}\right)  $. Furthermore if we consider that~$w_{m}\in
\lbrack0,1)$ then that point which describes an accelerated universe is always stable.
\end{itemize}

%

\begin{table}[tbp] \centering
\caption{Fixed points, cosmological parameters and stability for the
dynamical system  (\ref{L.11})-(\ref{L.11}) with exponential potential}%
\begin{tabular}
[c]{ccccccc}\hline\hline
\textbf{Point} & $\left(  \mathbf{x,y}\right)  $ & \textbf{Existence} &
$\mathbf{\Omega}_{m}$ & $\mathbf{w}_{tot}$ & \textbf{Acceleration} &
\textbf{Stability}\\\hline
$\mathit{P}_{A}$ & $\left(  1,0\right)  $ & $\lambda\in%
\mathbb{R}
$ & $0$ & $1$ & No & Unstable\\
$\mathit{P}_{B}$ & $\left(  \frac{3}{\lambda}\left(  1+w_{m}\right)  ,\frac
{3}{2\lambda}\left(  1-w_{m}\right)  \right)  $ & $\lambda\geq\frac{\left(
9+3w\right)  }{2}$ & $1-\frac{3\left(  3+w_{m}\right)  }{2\lambda}$ & $w_{m}$
& $~w_{m}<-\frac{1}{3}$ & $\lambda>\lambda_{1}^{B},~$\\
$\mathit{P}_{C}$ & $\left(  -\frac{6}{\lambda}+2,\frac{6}{\lambda}-1\right)  $
& $\lambda\in%
\mathbb{R}
^{\ast}$ & $0$ & $\frac{1}{3}\left(  2\lambda-9\right)  $ & $\lambda<\frac
{9}{2}$ & $\lambda<\frac{3}{2}\left(  3+w\right)  $\\\hline\hline
\end{tabular}
\label{tabl01}%
\end{table}%

The fixed points and the values of the physical variables on these points are
given in Table \ref{tabl01}.

\begin{figure}[ptb]
\includegraphics[height=7cm]{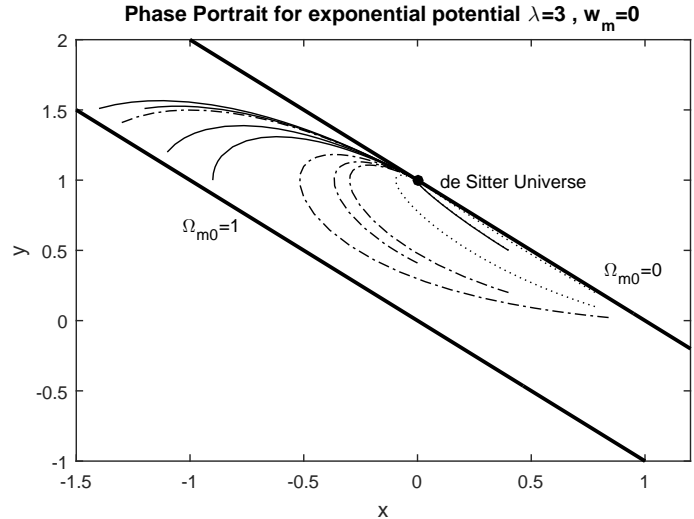}
\caption{Phase portrait for the potential $V\left(  \phi\right)  =\exp\left(
-\lambda\phi\right)  $~with $\lambda=-3$ and $w_{m}=0$. The stable point is a
de Sitter point. The left and right thick lines $y=1-x$, and $y=-x$, are the
borders \ in which $0\leq\Omega_{m}\leq1.$ The different lines describe
different initial conditions $\left(  x_{0},y_{0}\right)  $. }%
\label{plot1}%
\end{figure}

\begin{figure}[ptb]
\includegraphics[height=7cm]{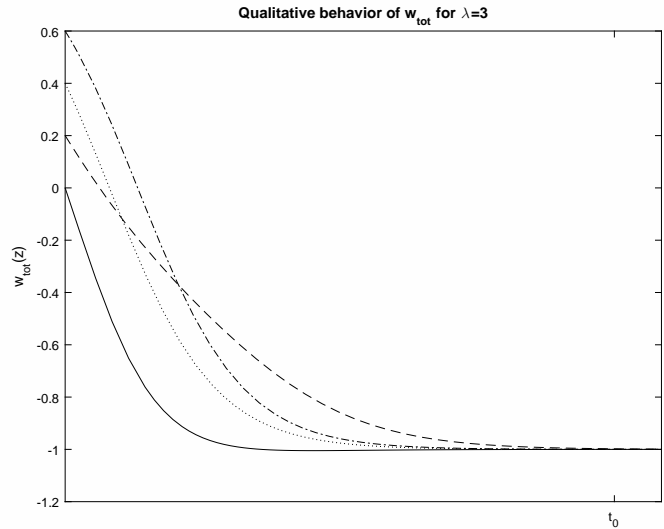}
\caption{Qualitative evolution of the total equation of state parameter
$w_{tot}$ for different initial conditions for the potential $V\left(
\phi\right)  =\exp\left(  -\lambda\phi\right)  $~with $\lambda=3$ and
$w_{m}=0$. The solid line is for initial conditions $\left(  x_{0}%
,y_{0}\right)  =\left(  0.4,0.5\right)  ,$ the dash-dash line for $\left(
x_{0},y_{0}\right)  =\left(  0.6,0.4\right)  ,$ the dot-dot line for $\left(
x_{0},y_{0}\right)  =\left(  0.7,0.3\right)  $ and the dash-dot line for
initial conditions $\left(  x_{0},y_{0}\right)  =\left(  0.8,0.2\right)  \,.$}%
\label{plot2}%
\end{figure}

\subsection{General potential}

Consider now a general potential $V\left(  \phi\right)  $, which corresponds
to a general function $F\left(  B\right)  $ and in a general function
$\bar{\Gamma}\left(  \lambda\right)  $. Now, if there exists a value
$\lambda=\lambda_{\ast}$ such as $\bar{\Gamma}\left(  \lambda\right)  =0$,
then from (\ref{L.09})-(\ref{L.10}) we find the fixed points $\bar{P}%
_{A},~\bar{P}_{B}$ and $\bar{P}_{C}$. The cosmological variables are the same
as those of Table \ref{tabl01}. However, the stability analysis is different.
There are two more possibilities for which the system (\ref{L.09}%
)-(\ref{L.11}) admit stationary points, $x=0$, or $\lambda=0$, with~$\lambda
^{2}\bar{\Gamma}\left(  \lambda\right)  $ well defined.

For $x=0$, we find that the rhs of (\ref{L.09})-(\ref{L.10}) vanishes at the
point $P_{D}$ with coordinates $\left(  x,y,\lambda\right)  =\left(
0,1,3\right)  $ while for $\lambda=0$ the fixed points are $P_{E}~,~\left(
x,y,\lambda\right)  =\left(  1,0,0\right)  $, where the latter is a special
case of the point $\bar{P}_{A}$ for $\lambda=0$. As far as concerns the
physical quantities at the point $P_{D}$ we have that $\Omega_{m}=0$ and
$w_{tot}=-1$, which means that $P_{D}$ is a de Sitter point. The stability of
the points it follows

\begin{itemize}
\item The eigenvalues of the linearized system around the point $\bar{P}_{A}$
are%
\begin{equation}
\bar{e}_{1}^{A}=3\left(  1-w_{m}\right)  ~,~\bar{e}_{2}^{A}=6-\lambda
_{0}~,~\bar{e}_{3}^{A}=-\lambda_{0}^{2}\bar{\Gamma}_{,\lambda}\left(
\lambda_{0}\right)
\end{equation}
from which we can see that $\bar{e}_{1}^{A}$ is always positive for $w_{m}%
\in\lbrack-1,1)$. Hence the point is always unstable.

\item For the point $\bar{P}_{B}$ the eigenvalues of the linearized system are%
\begin{equation}
\bar{e}_{1}^{B}=\frac{1}{4}\left[  -3\left(  1-w_{m}\right)  -\sqrt
{3\Delta_{B}}\right]  ~~,~e_{2}^{\bar{B}}=\frac{1}{4}\left[  -3\left(
1-w_{m}\right)  +\sqrt{3\Delta_{B}}\right]  ~
\end{equation}
and%
\begin{equation}
\bar{e}_{3}^{B}=-3\left(  1+w_{m}\right)  \lambda_{0}\bar{\Gamma}_{,\lambda
}\left(  \lambda_{0}\right)  ,
\end{equation}
where
\begin{equation}
\Delta_{B}=\left(  1-w\right)  \left(  75+21w_{m}-16\lambda_{0}\right)  .
\end{equation}
Eigenvalue $e_{3}^{B}$ is negative only when $\lambda_{0}\bar{\Gamma
}_{,\lambda}\left(  \lambda_{0}\right)  >0$. Now, for $\Delta_{B}\leq0$ we
have that $\mathit{Re}\left(  \bar{e}_{1}^{B}\right)  =\mathit{Re}\left(
\bar{e}_{2}^{B}\right)  <0$, which means that the point is always stable. On
the other hand for $\Delta_{B}>0$, $\bar{e}_{1}^{B}<0$ holds and the point is
stable when $e_{1}^{B}e_{2}^{B}>0$, that gives
\begin{equation}
\frac{3}{2}\left(  1-w_{m}\right)  \left(  9+3w_{m}-2\lambda_{0}\right)  <0,
\end{equation}
from which we find that $\bar{P}_{B}$ is stable when $\lambda_{0}>\frac{3}%
{2}\left(  3+w_{m}\right)  $.

\item For the point $\bar{P}_{C}$ we find the eigenvalues%
\begin{equation}
\bar{e}_{1}^{C}=-6+\lambda_{0}~,~\bar{e}_{2}^{C}=-9-3w_{m}+2\lambda_{0}%
~,~\bar{e}_{3}^{C}=-2\left(  -3+\lambda_{0}\right)  \bar{\Gamma}_{,\lambda
}\left(  \lambda_{0}\right)
\end{equation}
from which the point is stable when $\lambda_{0}<\frac{3}{2}\left(
3+w_{m}\right)  $ and $\left(  -3+\lambda_{0}\right)  \bar{\Gamma}_{,\lambda
}\left(  \lambda_{0}\right)  >0$.

\item At the point $P_{D}$ the matrix of the linearized system has the
following eigenvalues%
\begin{equation}
e_{1}^{D}=\frac{3}{2}\left(  -1-\sqrt{1-8\bar{\Gamma}\left(  3\right)
}\right)  ,~e_{2}^{D}=\frac{3}{2}\left(  -1+\sqrt{1-8\bar{\Gamma}\left(
3\right)  }\right)  ~,~e_{3}^{C}=-3\left(  1+w_{m}\right)
\end{equation}
which is a stable de Sitter point when $\mathit{Re}\left(  \bar{\Gamma}\left[
3\right]  \right)  >0$. Note that $P_{D}$ is a special point of $P_{C}$ when
$\lambda_{0}=3$. However, the eigenvalues of the linearized system are
different. That means that in a model with running $\lambda$, the two points
$P_{C}$ and $P_{D}$ can exist.

\item Finally the last point $P_{E}$ provides always a positive eigenvalue,
that is, the point is always unstable.
\end{itemize}

We conclude that for a general potential a second stable de Sitter point
exists which is stable for potentials in which $\bar{\Gamma}\left(  3\right)
>0$. In general two de Sitter phases are possible, the points $P_{C}$ and
$P_{D}$. The above results are collected in Tables \ref{table2} and
\ref{table3}.%

\begin{table}[tbp] \centering
\caption{Fixed points and cosmological parameters  for the
dynamical system  (\ref{L.11})-(\ref{L.11}) with arbitrary potential}%
\begin{tabular}
[c]{cccccc}\hline\hline
\textbf{Point} & $\left(  \mathbf{x,y,\lambda}\right)  $ & \textbf{Existence}
& $\mathbf{\Omega}_{m}$ & $\mathbf{w}_{tot}$ & \textbf{Acceleration}\\\hline
$\mathit{\bar{P}}_{A}$ & $\left(  1,0,\lambda_{0}\right)  $ & $\lambda_{0}\in%
\mathbb{R}
~,~\bar{\Gamma}\left(  \lambda_{0}\right)  =0$ & $0$ & $+1$ & No\\
$\mathit{\bar{P}}_{B}$ & $\left(  \frac{3}{\lambda}\left(  1+w_{m}\right)
,\frac{3}{2\lambda}\left(  1-w_{m}\right)  ,\lambda_{0}\right)  $ &
$\lambda_{0}\geq\frac{\left(  9+3w\right)  }{2}~,\bar{\Gamma}\left(
\lambda_{0}\right)  =0$ & $1-\frac{3\left(  3+w_{m}\right)  }{2\lambda}$ &
$w_{m}$ & $~w_{m}<-\frac{1}{3}$\\
$\mathit{P}_{C}$ & $\left(  -\frac{6}{\lambda}+2,\frac{6}{\lambda}%
-1,\lambda_{0}\right)  $ & $\lambda_{0}\in%
\mathbb{R}
^{\ast}~,~\bar{\Gamma}\left(  \lambda_{0}\right)  =0$ & $0$ & $\frac{1}%
{3}\left(  2\lambda-9\right)  $ & $\lambda_{\ast}<\frac{9}{2}$\\
$\mathit{P}_{D}$ & $\left(  0,1,3\right)  $ & Always & $0$ & $-1$ & Yes\\
$\mathit{P}_{E}$ & $\left(  1,0,0\right)  $ & Always & $0$ & $+1$ &
No\\\hline\hline
\end{tabular}
\label{table2}%
\end{table}%
%

\begin{table}[tbp] \centering
\caption{Eigenvalues and stability for the critical points of the
dynamical system  (\ref{L.11})-(\ref{L.11}) with arbitrary potential}%
\begin{tabular}
[c]{ccccc}\hline\hline
\textbf{Point/Eigenv.} & $\mathbf{e}_{1}$ & $\mathbf{e}_{2}$ & $\mathbf{e}%
_{3}$ & \textbf{Stability}\\\hline
$\mathit{\bar{P}}_{A}$ & $3\left(  1-w_{m}\right)  $ & $6-\lambda_{0}$ &
$-\lambda_{0}^{2}\bar{\Gamma}_{,\lambda}\left(  \lambda_{0}\right)  $ &
Unstable\\
$\mathit{\bar{P}}_{B}$ & $\frac{1}{4}\left[  -3\left(  1-w_{m}\right)
-\sqrt{3\Delta_{B}}\right]  $ & $\frac{1}{4}\left[  -3\left(  1-w_{m}\right)
+\sqrt{3\Delta_{B}}\right]  $ & $-3\left(  1+w_{m}\right)  \lambda_{0}%
\bar{\Gamma}_{,\lambda}\left(  \lambda_{0}\right)  $ & $\lambda_{0}>\frac
{3}{2}\left(  3+w_{m}\right)  $\\
$\mathit{P}_{C}$ & $-6+\lambda_{0}$ & $-9-3w_{m}+2\lambda_{0}$ & $-2\left(
-3+\lambda_{0}\right)  \bar{\Gamma}_{,\lambda}\left(  \lambda_{0}\right)  $ &
$%
\begin{array}
[c]{c}%
\lambda_{0}<\frac{3}{2}\left(  3+w_{m}\right)  ,\\
\left(  -3+\lambda_{0}\right)  \bar{\Gamma}_{,\lambda}\left(  \lambda
_{0}\right)  >0
\end{array}
$\\
$\mathit{P}_{D}$ & $\frac{3}{2}\left(  -1-\sqrt{1-8\bar{\Gamma}\left(
3\right)  }\right)  $ & $\frac{3}{2}\left(  -1+\sqrt{1-8\bar{\Gamma}\left(
3\right)  }\right)  $ & $-3\left(  1+w_{m}\right)  $ & $\mathit{Re}\left(
\bar{\Gamma}\left[  3\right]  \right)  >0$\\
$\mathit{P}_{E}$ & $3\left(  1-w_{m}\right)  $ & $6$ & $0$ &
Unstable\\\hline\hline
\end{tabular}
\label{table3}%
\end{table}%

As a special example consider the potential $V\left(  \phi\right)
=V_{0}e^{-\sigma\phi}+V_{1}$, from which we have%
\begin{equation}
F\left(  B\right)  =-\frac{B}{\lambda}\left(  \ln\left(  -\frac{B}{\lambda
}\right)  -1\right)  +V_{1}\text{.}%
\end{equation}

For that potential we have that $\phi=-\frac{1}{\sigma}\ln\left(
\frac{\lambda\bar{V}_{0}}{\sigma-\lambda}\right)  ,~\bar{V}_{0}=\frac{V_{1}%
}{V_{0}}$ and $\bar{\Gamma}\left(  \lambda\right)  =-1-\frac{1}{\bar{V}_{0}%
}\left(  1-\frac{\sigma}{\lambda}\right)  $. For the point $P_{C}$ we find
that $\lambda_{0}=\frac{\sigma}{1+\bar{V}_{0}}$. Hence $P_{C}$ is stable when%
\begin{equation}
\lambda_{0}<\frac{3}{2}\left(  3+w_{m}\right)  ~~\text{and~}-\left(
-3+\lambda_{0}\right)  \frac{\sigma}{\lambda_{0}^{2}\bar{V}_{0}}>0.
\end{equation}
Hence, if $3<\lambda_{0}<$ $\frac{3}{2}\left(  3+w_{m}\right)  $, then
$\frac{\sigma}{\bar{V}_{0}}<0$ while for $\lambda_{0}<3~$ the point is stable
when $\frac{\sigma}{V_{0}}>0$. On the other hand point $P_{D}$ is stable when
\begin{align}
\sigma &  <-3\left(  1+\bar{V}_{0}\right)  ~\ \text{for }\bar{V}_{0}>0,\\
\sigma &  >-3\left(  1+\bar{V}_{0}\right)  ~\text{~for }\bar{V}_{0}<0.
\end{align}

In the following section we proceed with the derivation of some analytical
solutions for the field equations (\ref{L.03})-(\ref{L.06}).

\section{ Exact cosmological solutions}

\label{solutions}

We consider that in the field equations (\ref{L.03})-(\ref{L.06}) the matter
source corresponds to that of a dust fluid, i.e. $w_{m}=0$, and $p_{m}=0$.
Hence (\ref{L.06}) provides $\rho_{m}=\rho_{m0}a^{-3}$. \ Furthermore for the
potential, $V\left(  \phi\right)  $, we consider that $V_{1}\left(
\phi\right)  =V_{0}\exp\left(  -3\phi\right)  $, which leads to a de Sitter
universe and $V_{2}\left(  \phi\right)  =V_{0}\exp\left(  -3\phi\right)
-2\Lambda,~$. According to the above this has two de Sitter phases, points
$P_{C}$ and $P_{D}$.

\subsection{Solution for $V\left(  \phi\right)  =V_{0}\exp\left(
-3\phi\right)  $}

For the potential $V_{1}\left(  \phi\right)  $ the Lagrangian of the field
equations becomes%
\begin{equation}
\mathcal{L}\left(  a,\dot{a},\phi,\dot{\phi}\right)  =-6a\dot{a}^{2}%
+6a^{2}\dot{a}\dot{\phi}\,-a^{3}V_{0}e^{-3\phi} \label{tl.01}%
\end{equation}
so that the field equations are the Euler-Lagrange equations of (\ref{tl.01})
with respect to the variables $\left\{  a,\phi\right\}  ,$ while the first
modified Friedmann's equations can be seen as the Hamiltonian function of
(\ref{tl.01}),$~\mathcal{H}=E,$ where now $\rho_{m0}=2\left\vert E\right\vert
.$

It is straightforward to see that (\ref{tl.01}) admits the two extra
Noetherian conservation law\footnote{For the application of point symmetries
in cosmological studies see \cite{ref2aaa,ref2,ref4} and references therein
while a partial classification of Noether point symmetries in $f\left(
T,B\right)  $ can be found in \cite{ref2c}} which are%
\begin{equation}
I_{1}=\dot{\phi}-\frac{\dot{a}}{a}~\mbox{\rm and}~I_{2}=t\left(  \dot{\phi
}-\frac{\dot{a}}{a}\right)  -\left(  \phi-\ln a\right)  . \label{tl.02}%
\end{equation}

We perform the coordinate transformation $a=u^{\frac{1}{3}},~\phi=v-\frac
{1}{3}\ln\left(  u\right)  $. In the new coordinates Lagrangian (\ref{tl.01})
is written%
\begin{equation}
L\left(  u,\dot{u},v,\dot{v}\right)  =2\dot{u}\dot{v}-V_{0}e^{-3v}
\label{tl.03}%
\end{equation}
and the field equations are taking the simple form
\begin{equation}
2\dot{u}\dot{v}+V_{0}e^{-3v}=2E,
\end{equation}%
\begin{equation}
\ddot{u}-\frac{3}{2}V_{0}e^{-3v}=0~\mbox{\rm and}~\ddot{v}=0.
\end{equation}

Finally the solution is given in a closed-form expression as follows%
\begin{equation}
u\left(  t\right)  =\frac{\bar{V}_{0}}{9v_{1}^{2}}e^{-3v_{1}t}+u_{1}t+u_{0},
\label{tl.04}%
\end{equation}
where $\bar{V}_{0}=\frac{3}{2}V_{0}e^{-3v_{0}}$ and $E=u_{1}v_{1}$. We have
that the de Sitter phase is recovered when $v_{1}<0$. \ From (\ref{tl.04}) it
follows that the scale factor has the form%
\begin{equation}
a^{3}\left(  t\right)  =a_{2}e^{\beta t}+a_{1}t+a_{0},
\end{equation}
where the spacetime has a singularity at $t=0$ when $a_{0}=-a_{2}$, that is,
the scale factor becomes $a^{3}\left(  t\right)  =a_{2}\left(  e^{\beta
t}-1\right)  +a_{1}t$.

Moreover, in the vacuum solution in which $u_{1}v_{1}=0$, we have two
possibilities: $u_{1}=0$ or $v_{1}=0$. \ For the latter case, that is
$\beta=0$, the solution is $a\left(  t\right)  \simeq t^{\frac{1}{3}}$, which
corresponds to the solution of GR with a perfect fluid with equation of state
parameter $w=-\frac{1}{3}$.

However, in the latter case for which $u_{1}=0$ i.e., $a_{1}=0$, the scale
factor with $a\left(  t\rightarrow0\right)  =0$ is of the form
\begin{equation}
a^{3}\left(  t\right)  =a_{2}\left(  e^{\beta t}-1\right)  .
\end{equation}
Easily we have that $t=\frac{1}{\beta}\ln\left(  1+\frac{a^{3}}{a_{2}}\right)
$, from which we calculate the Hubble Function%
\begin{equation}
\left(  H\left(  a\right)  \right)  ^{2}=\frac{\beta^{2}}{9}+\frac{2\beta^{2}%
}{9}a_{2}a^{-3}+\frac{\beta^{2}}{9}\left(  a_{2}\right)  ^{2}a^{-6}.
\end{equation}
This means that the theory provides us with a cosmological constant term, a
dust term and a stiff fluid, equivalently with that of the minimally coupled
scalar field \cite{anbar}.

\subsection{Solution for $V\left(  \phi\right)  =V_{0}\exp\left(
-3\phi\right)  -2\Lambda$}

As a second potential we consider the same as the above where now we include a
cosmological constant term. It is easy to see that in the coordinate system
$\left\{  u,v\right\}  $ the field equations become%
\begin{equation}
2\dot{u}\dot{v}+V_{0}e^{-3v}-2\Lambda u=2E
\end{equation}
and
\begin{equation}
\ddot{u}-\frac{3}{2}V_{0}e^{-3v}=0~,~\ddot{v}-\Lambda=0,
\end{equation}
from which we have that the scale factor is expressed in terms of the error
function, $\mathcal{E}\left(  t\right)  $, as%
\begin{align}
a^{3}\left(  t\right)   &  =\frac{\bar{V}_{1}}{3\Lambda}\exp\left(  -\frac
{3}{2}\Lambda t^{2}-3v_{1}t\right)  +\\
&  +\frac{\sqrt{6\pi}\bar{V}_{1}}{6\Lambda^{\frac{3}{2}}}\left(  \Lambda
t+v_{1}\right)  \mathcal{E}\left(  \frac{\sqrt{6}}{2\Lambda}\left(  \Lambda
t+v_{1}\right)  \right)  +u_{1}t+u_{2},
\end{align}
where $\bar{V}_{1}=V_{0}e^{-3v_{0}}$ and $v\left(  t\right)  =\frac{\Lambda
}{2}t^{2}+v_{1}t+v_{0}$.

The reason that this is possible is that the Lagrangian of the field equations
admits a Noetherian conservation law which is not generated by point
symmetries as in the potential $V_{1}\left(  \phi\right)  $ but from
generalized symmetries. In particular the Killing tensor of the minisuperspace
provides a contact symmetry (see \cite{palfr} and references therein).

\section{Conclusions}

\label{conclusions}

In the context of modified theory of gravities we considered a gravitational
theory in which the deviation from General Relativity is given by a function
of the boundary term which relates the Ricci Scalar, $R$, and the invariant,
$T$, of teleparallel gravity. The theory that we considered is a fourth-order
theory and in the case of an isotropic and homogeneous universe the field
equations can be written as a (constraint) Hamiltonian system with two degrees
of freedom. One degree of freedom corresponds to the scalar factor of the
geometry and the second one is a field which describes the higher-order
derivatives, as in the case of $f\left(  R\right)  $-gravity. The theory
admits a constraint and it is the equation of motion which corresponds to the
lapse function of the geometry.

Though that theory is a fourth-order theory differs from $f\left(  R\right)
$-gravity and the field which is introduced from the application of the
Lagrange multipliers does not describe a scalar tensor theory.\ The
minisuperspace Lagrangian is given by%

\begin{equation}
\mathcal{L}\left(  a,\dot{a},\phi,\dot{\phi}\right)  =-\frac{6}{N}a\dot{a}%
^{2}+\frac{6}{N}a^{2}\dot{a}\dot{\phi}+Na^{3}V\left(  \phi\right)  .
\label{ln001}%
\end{equation}
However, under the change $a=Ae^{\frac{\phi}{2}}$ and $N=e^{\frac{3\phi}{2}}%
n$, this Lagrangian becomes
\begin{equation}
\mathcal{L}\left(  A,\dot{A},\phi,\dot{\phi}\right)  =-\frac{6}{n}A\dot{A}%
^{2}+\frac{3}{2n}A^{3}\dot{\phi}^{2}+na^{3}\left(  e^{3\phi}V\left(
\phi\right)  \right)  \label{ln002}%
\end{equation}
which is the Lagrangian of a canonical minimally coupled (phantom) scalar
field with potential $U\left(  \phi\right)  =e^{3\phi}V\left(  \phi\right)  $.
It is easy to see that the transformation $\left(  N,a\right)  \rightarrow
\left(  e^{\frac{3\phi}{2}}n,Ae^{\frac{\phi}{2}}\right)  $ does not relate
conformal equivalent theories, such as in the scalar-tensor theories. However,
the relation between the two Lagrangians, (\ref{ln001}) and (\ref{ln002}), is
important because the analysis of \cite{ssn1} can be applied and it can be
easily shown that the gravitational field equations (\ref{L.03})-(\ref{L.05})
form an integrable dynamical system.

In order to study the effects which follow from the new terms in the dynamics
of the field equations the critical points were calculated. Every point
corresponds to a physical state and the physical parameters were calculated.
The importance of the existence of the points is that for families of initial
conditions the evolution of the universe passes closely to the physical states
which are described from the points\ (unstable points) or at the end reach the
solution which is described by the critical point (stable point). In our
analysis we found that for it is possible to have a theory which provides a
matter era (unstable point) and two acceleration phases in which the one can
be stable and the other unstable. \ This is an interesting result and it is
different from that of $f\left(  R\right)  $-gravity. Moreover some
closed-form solutions were derived and the explicitly form of the FLRW
spacetime was found.

There are various open questions which have to be answered for that
consideration, but the property that the only constraint in the field
equations is that of the \textquotedblleft Hamiltonian\textquotedblright\ is
essential because various methods can be applied, from the scalar field
description, in order to study the theory. In a future work we would like to
extend the present analysis in order to search for other kinds of cosmological
solutions and extend the analysis of the critical points at the infinite
region. The existence of static-spherical solution is also of special interests.

\begin{acknowledgments}
I acknowledge the financial support of FONDECYT grant no. 3160121 and I thank
the Durban University of Technology and the University of KwaZulu-Natal for
the hospitality provided while part of this work was performed.
\end{acknowledgments}

\end{document}